\documentstyle[12pt,epsf]{article}
\begin{document}
%
% Enter the title
%
\begin{center}
{\Large Criticality and Punctuated
Equilibrium in a Spin System Model of a Financial Market}
\end{center}
%
% Enter your name
%
\begin{center}
A.~PONZI and Y.~AIZAWA  
\end{center}
\begin{center}
Department of Applied Physics, School of Science and Engineering,
Waseda University, Shinjuku, Tokyo 169-8555, Japan\\
\end{center}
\baselineskip=32pt
\noindent
{\Large \bf Abstract}
%
% write the abstruct
%

%
% write your manuscript
%
We describe a financial market model which shows a non-equilibrium
phase transition. Near the transition punctuated equilibrium behaviour
is seen, with avalanches occuring on all scales. This scaling is
described by an exponent very near 1. This system shows intermittent
time development with bursts of global synchronization reminiscent of
a market rollercoaster.

\section{Introduction}

Recently much attention has been given to connections between
criticality and self-organized criticality (SOC) and evolutionary
phenomenon, particularly punctuated equilibrium, and to
connections between SOC and synchronization. 
We describe a model 
which we hope draws some connection between these 2 ideas.
SOC has been proposed to describe out of equilibrium systems
that are critical, that self-organize into a scale invariant
critical state without tuning of a control
parameter and show fractal time series.\cite{rf:52} 
Evolution SOC type models \cite{rf:60,rf:68,rf:45}
have been proposed to explain 
punctuated equilibrium. \cite{rf:66,rf:67}
Punctuated equilibrium is the phenomenon observed
in the fossil record where long periods of stasis are interrupted 
by sudden bursts of evolutionary change. 
Kaufman and Johnsen \cite{rf:64} have modeled co-evolution,
where agents live on 
a coupled fitness landscape and walk around by
random mutation, only 
moves to higher fitnesses are allowed.
Once at a local maximum the walk stops until moves by another
agent deform the lanscape so the agent is no longer at a maximum.
Kauffman \cite{rf:64} has linked this to SOC. Bak,
Sneppen and Flyvbjerg \cite{rf:60,rf:68} have taken a similar
approach. They define a species as 
a barrier to increasing fitness and choose the
least fit, then randomly change its barrier and the barriers of
other agents. The
system evolves to a critical state with a self-organized
fitness threshold.
SOC has also been linked to periodic
behaviour.\cite{rf:50,rf:56,rf:58,rf:70} 
A.Corral et al and Bottani \cite{rf:56,rf:50}, say there is a close relationship between
SOC and synchronization. SOC
appears when a system is perturbed which
otherwise should synchronise totally or partially. 
The perturbation may be open boundary conditions rather than
periodic\cite{rf:50,rf:70} or it may be
randomness present in initial conditions which is preserved by
the dynamic, or it may be the addition of noise.
The model we study appears to link these 2 ideas with the
emergence of avalanches of partial synchronization on all
scales. However all these models are real space models whereas
ours is a mean-field model with no spatial dimension. Our model 
is entirely determistic, the critical state is produced by
certain initial conditions, indeed other initial conditions
produce completely periodic states. 
Our model also is not strictly speaking an SOC model since
critical behaviour only occurs for a certain range of the
parameter and then only for certain initial conditions. A
complete analysis of the 
intial conditions is outside the scope of this paper.

Our model was originally
motivated as a model of the behaviour of speculating traders in a 
financial market in the spirit of co-evolution.
Recent results have shown stock price time
series to be fractal with Hurst exponent different from 0.5,
\cite{rf:72} 
and with positive lyapunov exponent.
\cite{rf:33,rf:34,rf:39}
Scrambling daily returns changes the Hurst exponent back to 0.5.  
Large crashes have been supposed
to be due to exogenous shocks, where information enters the
market randomly.
However large crashes interspersed with periods of slow
growth are 
strongly reminiscent of punctuated equilibrium. Indeed
Mandlebrot has noted large changes of cotton prices occur in
oscillatory groups 
and the movement in tranquil periods is smoother than
predicted.\cite{rf:77} 
Scaling behavior has been noted in a financial 
index and in the size of companies.\cite{rf:100,rf:101}  Stanley et
al have noted `scaling laws used to describe complex systems
comprised of many interacting inanimate particles (as in many
physical systems) may be usefully extended to describe complex
systems comprised of many interacting animate subsystems (as in
economics).'

Various models have been proposed to explain
market movements.\cite{rf:36,rf:38,rf:6}
Sato and Takayasu have proposed a threshold type
model.\cite{rf:73,rf:73a}
Since critical states can produce avalanches on all scales,
without the need for 
exogenous shocks, we believe critical type dynamics
are present in financial market dynamics. 

\section{Model}
\noindent
We hope to model co-evolutionary phenomenon where the micro-level
itself defines the macro-level but is also slaved to the
macro-level. This is very evident in speculative financial market dynamics
where a collection of individuals (micro) trade therby creating a price time
series (macro), but determine their trading behaviour by reference to
this same price series and other macro variables.  
We desired to make a model in analogy to this phenomenon. 

This is a highly stylized toy model of a stock-market.There are 
$N$ agents which are represented by spins $s_{i}(t)$, where
$s_{i}(t)=1$, means the agent $i$ owns the stock and
$s_{i}(t)=-1$ means doesn't own the stock at time t. Each agent also has an 
absolute fitness $F_{i}(t)$ and a relative fitnes
$f_{i}(t)=F_{i}(t)-F(t)$ where the mean-fitness
$F(t)=\frac{1}{N}\sum_{i=1}^{N}F_{i}(t)$. 
We
believe speculative traders are part of 2 crowds, bulls and bears, and our
macrovariable is `groupthink' $G(t)$, 
defined by, 
\begin{equation}
G(t)=\Delta P(t)=\frac{1}{N}\sum_{i=1}^{N}s_{i}(t)
\end{equation}
The dynamic is:
\begin{equation}
\Delta s_{i}(t)=s_{i}(t+1)-s_{i}(t)=\left\{\begin{array}{ll}
-2s_{i}(t) & f_{i}(t)\leq 0\\
0 & f_{i}(t)>0
\end{array}
\right.
\end{equation}
\begin{equation}
\Delta F_{i}(t)=F_{i}(t+1)-F_{i}(t)=-\frac{1}{2}\Delta s_{i}(t)G(t)+\frac{1}{2}|\Delta s_{i}(t)|c
\end{equation}

The dynamic is synchronous and deterministic. First $G(t)$ and
$F(t)$ are calculated then all agents are updated according to
(2) and (3). The price $P(t)$ is defined by (1) and
$P(0)=0$. Initially $s_{i}(0)$ are chosen randomly with
probablity 1/2 and $F_{i}(0)$ are chosen randomly from the
interval [-1,1].

$G(t)$ measures the bullishness or bearishness of the crowd.
Although different to ours Callan and Shapiro mention groupthink
in Theory of Social 
Imitation\cite{rf:75} and Vaga's\cite{rf:76,rf:72} Coherent
Market Hypothesis explicitly includes a variable called
{\it groupthink}. 

We believe speculative agents determine their spin state
dependent on whether they 
believe the market will move in their favour in the future. 
Therefore 
our agents have an absolute fitness $F_{i}(t)$ which measures their
perception of whether they are in a good position with respect
to the future. If $F_{i}(t)$ is relatively high their state will
be stable and if $F_{i}(t)$ is relatively low they will want to
change their current state. Many ways to define $F_{i}(t)$ are
possible. In this model we define it by 
analogy to Plummer. Agents
consider the market to be `overbought' or `oversold'. 
In our simplistic model this is measured by $G(t)$. An agent is fit if it is in the minority
group.  According to Plummer when most agents 
are in one position then there must be less buying into this
position (because there are only a finite amount of agents) and
therefore the market will eventually correct itself (change
direction) because 
its growth will not be sustainable. It is always profitable then 
to be in the minority group before a correction. At a correction 
the dominant crowd breaks, the macro position dissolves, the market
may crash, and 
subsequently bull and bear crowds will begin to reform. 
In fact at these times the agents may trade in
2 macro-clusters or chaotically with the market attaining high
volatility which persists for some time.  
This type of
trader has 
been called a {\it sheep trader} \cite{rf:36,rf:38}, in
contrast to fundamentalists speculators and
non-speculators. 
Therefore in this model an agents fitness is increased if it
changes from the 
majority group to the minority group, with the increase
proportional to the size of the majority. The opposite is
applied if it changes the other way. 
If an agent doesn't change its state then it's absolute fitness
$F_{i}(t)$ is not
changed, regardless of whether $G(t)$ changes.
An agent also has a relative fitness $f_{i}(t)$. The $f_{i}(t)$ are the
behaviour controlling variables in this model. They may change
in 2 ways. Firstly an agent $\alpha$ may change its state $s_{\alpha}(t)$ thereby
directly changing $F_{\alpha}(t)$ and $f_{\alpha}(t)$. This is similar to
a single adaptive
move on a fitness landscape by an individual optimizing
agent. Secondly co-evolution may occur. Here an agent's
relative fitness $f_{\beta}(t)$ may change due to changes in the
other agents fitnesses $F_{i}(t)$ changing $F(t)$ while
$F_{\beta}(t)$ remains constant.

To model evolution then we follow natural selection by analogy
and mutate unfit agents and leave fit agents unchanged (although 
their relative fitnesses may change). 
as in Kaufman,\cite{rf:64} and Bak et al\cite{rf:60,rf:68}.
Mutation is considered to
be a state change and this changes an agent's fitness according
to (3). In this model since
their are only 2 possible states this means we simply flip
state. (In a more extensive model this would correspond to
changes to an
ownership portfolio vector). To decide which spins flip we could
compare pairs of 
fitnesses and change the least fit. 
That is we could choose 2 agents $\alpha$ and $\beta$ and let
them compete so that
$F_{\alpha}>F_{\beta}$ then we say
$s_{\alpha}(t+1)=s_{\alpha}(t)$ and
$s_{\beta}(t+1)=-s_{\beta}(t)$. 
However in this paper we simply
take a mean-fitness approach. That is all agents $i$ whose
fitnesses $F_{i}(t)$ fulfill $F_{i}(t)\leq F(t)$ ie
$f_{i}(t)\leq 0$
flip there spins and their fitnesses change according to (3).
All other agent's states and absolute fitnesses $F_{i}(t)$ do
not change although their relative fitnesses $f_{i}(t)$ of
course do.  
Therefore fit agents which could be considered to be at a local
maximum do not change their states until the mean-fitness $F(t)$ has
become equal to their fitnesses $F_{i}(t)$.

This means the fitnesses are all internally defined emergent
properties as in co-evolution. Of course if there is no overall crowd
polarisation then changing state does not change fitness.

Therefore the fitness update rule (3) can be seen as the
adaptive walk part and this is the reason why we do not simply
set  $F_{i}(t)=-s_{i}(t)G(t)$ or $\Delta F_{i}(t)=-s_{i}(t)G(t)$
continuously for all agents. We 
hope agents will take time to walk out of unfit states and that
fit maxima will be created which persist for some time. Our
absolute fitness is therefore cumulative and is only changed for unfit
agents. More realistically we could
think of agents imperfectly sampling the market ie $G(t)$ at a series of
times to determine their current absolute fitness. In
fact we see that the concept of relative fitness and absolute
fitness are very similar to the concept of `bounded
rationality.' An agents rationality is bounded because he only
makes local adaptive moves and can percieve only his absolute
fitness but not the overall mean-fitness or his relative fitness

This fitness of the position is natural in the sense that it can be seen as 
a kind of potential for future profit, usually termed `utility'
in economics. The fitter an agent is
the less likely it will want to change its state, the more
stable it is, because it believes the market to be oversold in
its favour.

Since $G(t)$ will on average be $0$,
in addition in equation (3) we include a very small control
parameter $c$ which controls the driving rate. We add this to 
all fitnesses below the mean so that unfit agents on
average will be come fitter and interact with the fit
agents. This is a general characteristic of evolutionary systems 
that single entity moves on fitness landscape should be on
average uphill. 

Our market price $P(t)$ is defined by $\Delta P(t)=G(t)$, ie
price increases while more people own the share than don't own
it, and the price is theoretically unbounded as it should
be. Positive groupthink means positive increase and
vice-versa. This is similar to the way prices are usually
defined by $\Delta p(t)\propto Z(t)$, where $Z(t)$ 
is the excess demand for something. 
This model does not included a fixed amount of
shares. Indeed any trader may independently buy or sell a share
without the notion of swapping. This reflects the fact that this 
is a model of only speculative behaviour, and part of a much
larger of pool of shares. However a more realistic model should
include a fixed amount of shares.

This model is intended to be a suggestive
illustration rather than a realistic stock-market. 

\section{Results}

Shown in Fig.1a is a time series for the fitnesses $F_{i}(t)$
for an $N=80$ system for $c=0.01$. Punctuated equilibrium behaviour is
clearly visible, with periods of relative stasis interspersed
with sudden jumps.
Although not shown the mean-fitness time series $F(t)$
shows changes on all scales similar to a devils staircase. Also
shown in Fig.1b is the corresponding daily returns time-series
$\Delta P(t)=G(t)$,
this also shows calm periods and sudden bursts of high
volatility. Infact this behaviour is a kind of intermittent
partial synchronization. Shown in Fig.2 is the same time
series but with a small portion magnified. Macroscopic 
synchronization can be seen. Partial synchronizations show various
different periods and complexities, and persist for various
lengths of time.
Clustering allows
synchronized spins to trade in phase with groupthink $G(t)$ thereby
rapidly increasing there fitnesses, or out of phase thereby
becoming less fit, this is the origin of the sudden large
changes in fitness. 
Also at these times the fitness deviation suddenly
increases,(not shown). Between periods of large-scale partial
synchronization with high volatility, periods 
of calm are characterized by a small even number of spins
flipping in anti-phase, they therefore increase their $F_{i}$
only slowly due to the driving parameter c, the returns $G(t)$
remaining roughly 
constant during these periods with the fitness deviation
decreasing. (Of course anti-phase flipping with no average
increase in fitness is prevented in a real market by a
fixed transaction cost. A more realistic model must include this.) 
When the mean-fitness $F(t)$ which is usually increasing
crosses some non-flipping $F_{i}$, this spin flips and may cause the
$F(t)$ to cross some more $F_{i}$, possibly starting an 
avalanche. This only happens when the total fitness deviation is 
small.

Shown in Fig.3 are two price $P(t)$ time series for
$c=0.013$. Their fractal slightly repetetive pattern is highly
reminiscent of real financial time series.

Since this model is deterministic, completely periodic
states are also possible. Shown in Fig.4 is the average of the
quantity $<R(t)>$
where $R(t)=\sum_{i=1}^{N}|\Delta s_{i}(t)|$
is the amount of spins which flip at any time and $<\ldots >$
denotes time averaging. In Fig.4a are time series for $c=0.0113$ 
while Fig.4b is $c=0.01$. 

Fig.4a shows one time series finds
the 2 cluster periodic state, where 2 groups alternately topple. 
Here 59 time series are included in the non-periodic state. If
this is a
transient it is super long even for the moderate size 
$N=200$. Fig.4b shows at more regular $c$ values the system
can find periodic states with larger amounts of
clusters. Roughly half of the 60 series investigated become
periodic by $t=2.5\times 10^{7}$.
Shown in Fig.5a is $<R(t)>$ plotted against $c$. 
Infact to construct this plot we
discarded $3\times 10^{7}$ time steps and then averaged over the next
$20,000$, each point represents a different initial condition
and there are $8$ for each cost $c=0.001x+0.000138$, $x$ is an integer.
For small cost
critical type behaviour is evident 
with a sudden phase transition at $c=0$. This is of course
because at negative $c$ less fit spins continuously flip and
never interact with spins at greater than mean fitness. In fact
the system divides into a frozen solid fit component and an
unfit gaseous type component for negative $c$. The size of the
frozen component 
depends on the initial conditions as can be seen from the points 
at negative cost. For larger positive cost an upper branch of
periodic attractors at  
$<R(t)>=100$, half the system size, is evident, the
system has settled into 2 alternately toppling clusters which
interleave the mean-fitness $F(t)$. The lower branch is
characterised by the punctuated equilibrium 
state shown in Fig.1.
Fig.5b shows the time average $<S(t)>$ of an entropy type
quantity of the fitness
distribution $S(t)$ given by,
$S(t)=-\sum_{i=1}^{N}\frac{|f_{i}(t)|}{f(t)}ln\frac{|f_{i}(t)|}{f(t)}$
where $f_{i}(t)$ is the fitness deviation and 
$f(t)=\sum_{i=1}^{N}|f_{i}(t)|$. The averaging is the same as
for $<R(t)>$. 
For this $N=200$ system the maximum $S=ln200\approx
5.3$ and the periodic points at positive and negative cost are
very near this. 
The punctuated equilibrium state which exists near
the transition is more ordered at lower entropy. 

This is our first evidence of critical behaviour for small $c$. 
Second evidence is obtained by 
looking at the distribution of avalanches. In the punctuated
equilibrium state the system finds a state characterised by
fluctuations on all scales. Shown in Fig.6 is the
distribution of $P(R)$ against $R$ where $P(R)$ is the
probability of an avalanche of size $R$. These are distributions 
of avalanches for 1 time series for 3 different system
sizes. They are not ensembles of time series, this distribution
is independent of the initial conditions and any non-periodic
time series contains all avalanche sizes. 

The time series
were of length $T=16,000,000$, near the transition point at
$c=0.0113$. 
The distribution shows scale invariance,$P(R)\sim R^{\alpha }$ up
to about 
half the system size. At half the system size there is a peak
where the system almost finds the periodic attractor and spends
more time in these states.
After this 
the distribution continues  
to the cutoff near the system size. The scaling exponent $\alpha 
$ taken
from the $N=3000$ distribution is $\alpha =-1.085\pm 0.002$.
Also shown in Fig.7 is the distribution of magnitude of
changes in mean 
fitness $\Delta F(t)=|F(t+1)-F(t)|$, the steps in the devils
staircase. 
The time series are
the same as in Fig.6 for 2 different system sizes, there is
no ensemble averaging. In fact the two distributions for
$N=1500,923$ are almost identical, if we were to superpose them,
only one could be seen. This is also true for other system
sizes. Peaks appear at $\Delta F
\approx 0.12,0.5,0.75$. Between the peaks we see scaling
regimes. Here we see at least 2
scaling regimes, $P(\Delta F)\sim \Delta F^{\beta}$ where for
$\Delta F \le 0.1$,  $\beta =-1.25 \pm 0.003$ and for $0.1\le
\Delta F \le 0.4$, $\beta =-1.39 \pm 0.02$. Possibly there is
another scaling regime for $0.55\le \Delta F \le 0.7$.
\section{Conclusion}
\noindent
This model illustrates an interesting relation between critical
phenomenon and punctuated equilibrium on the one hand, and
between partial synchronization and punctuated equilibrium on
the other hand. The system synchronizes for certain cost $c$ and
certain intial conditions, otherwise it shows critical
behaviour, similar to the SOC models cited in the introduction. We
believe this deserves further  
investigation. We also find an interesting phase transition. 

Some typical behaviour of 
money markets is present here, especially the periods of low
volatility, where the price is relatively stable and the fitness 
grows slowly while the fitness deviation decreases slowly, 
interrupted by shorter periods of persistent high volatility and macroscopic
oscilations which are observed in real time series. We wonder if 
like in earthquake dynamics, which are often modelled by SOC
dynamics, a large crash in a real financial market is preceded
by some smaller self-reinforcing oscilatory pre-shock, as is
seen in our dynamics here. Also the time price time series is
highly suggestive of real time series, with 
formations similar to `double tops' and `rebounds' described in
quantitative analysis, produced by the 
near periodic macro behaviour which can appear. The
slightly repetetive self-similarity reminds us of financial time series.    

Many possible models of financial market dynamics can be
plausibly suggested, included many exhibiting threshold
dynamics, since data concerning the micro behaviour of
individual traders is not available.

\end{document}